\documentstyle[epsfig]{aipproc}

\def\eg{{e.g., }}
\def\ie{{i.e., }}
\def\etal{{et al., }}
\def\et{{et al. }}
\def\etc{{etc.}}

\def\'{^{\prime}}

\def\hmpc{{\, {\rm h}^{-1}~\rm Mpc}}

\def\kms{{\rm~km~s^{-1}}}

\def\kpc{{\rm~kpc}}
\def\mpc{{\rm~Mpc}}

\def\spose#1{\hbox to 0pt{#1\hss}}
\def\lta{\mathrel{\spose{\lower 3pt\hbox{$\mathchar"218$}}
     \raise 2.0pt\hbox{$\mathchar"13C$}}}
\def\gta{\mathrel{\spose{\lower 3pt\hbox{$\mathchar"218$}}
     \raise 2.0pt\hbox{$\mathchar"13E$}}}
\def\ge{\mathrel{\spose{\lower 3pt\hbox{$-$}}
     \raise 2.0pt\hbox{$\mathchar"13E$}}}
\def\le{\mathrel{\spose{\lower 3pt\hbox{$-$}}
     \raise 2.0pt\hbox{$\mathchar"13C$}}}

\begin{document}

\title{{The Quintessential CMB, Past \& Future}}

\author{J. Richard Bond$^{1}$, Dmitry Pogosyan$^{1}$, Simon Prunet$^{1}$, Kris Sigurdson$^{1}$ and
        the MaxiBoom Collaboration$^{2}$}
\address{1. CIAR Cosmology Program, Canadian Institute
for Theoretical Astrophysics, \\
        60 St. George St., Toronto, ON M5S 3H8, Canada \\
      2. See Jaffe \et 2000 \cite{jaffe00} for the full author and institution
        list.\\
       {\small \tt CITA-2000-64, in Proc. CAPP-2000 (AIP), eds.  R. Durrer, J.
       Garcia-Bellida, M. Shoposhnikov \normalsize}
        }

\maketitle

\begin{abstract}
The past, present and future of cosmic microwave background (CMB)
anisotropy research is discussed, with emphasis on the Boomerang
and Maxima balloon experiments. These data are combined with large
scale structure (LSS) information derived from local cluster
abundances and galaxy clustering  and high redshift supernova
(SN1)  observations to explore the inflation-based cosmic
structure formation paradigm. Here we primarily focus on a
simplified inflation parameter set,
$\{\omega_b,\omega_{cdm},\Omega_{tot}, \Omega_Q,w_Q,n_s,\tau_C,
\sigma_8\}$. After marginalizing over the other cosmic and
experimental  variables, we find the current CMB+LSS+SN1 data
gives $\Omega_{tot} = 1.04\pm 0.05$, consistent with (non-baroque)
inflation theory. Restricting to $\Omega_{tot}=1$, we find a
nearly scale invariant spectrum, $n_s =1.03 \pm 0.07$. The CDM
density, $\omega_{cdm}=0.17 \pm 0.02$, is in the expected range,
but the baryon density, $\omega_b\equiv \Omega_b {\rm h}^2 =
0.030\pm 0.004$, is slightly larger than the current $0.019\pm
0.002$ Big Bang Nucleosynthesis estimate. Substantial dark
(unclustered) energy is inferred, $\Omega_Q \approx 0.68 \pm
0.05$, and CMB+LSS $\Omega_Q$ values are compatible with the
independent SN1 estimates. The dark energy equation of state,
parameterized by a quintessence-field pressure-to-density ratio
$w_Q$, is not well determined by CMB+LSS ($w_Q < -0.3$ at 95\%
CL), but when combined with SN1 the resulting $w_Q < -0.7$ limit
is quite consistent with the $w_Q$=$-1$ cosmological constant
case. Though forecasts of statistical errors on parameters for
current and future experiments are rosy, rooting out systematic
errors will define the true progress. \vspace{1pc}
\end{abstract}

\section*{CMB Analysis: Past, Present and Future}

The CMB is a nearly perfect blackbody of $2.725 \pm 0.002\, K$
\cite{matherTcmb}, with a $3.372 \pm 0.007\, mK$ dipole associated
with the 300 $\kms$ flow of the earth in the CMB, and a rich
pattern of  higher multipole anisotropies at tens of $\mu$K
arising from fluctuations at photon decoupling and later. Spectral
distortions from the blackbody associated with starbursting
galaxies detected in the COBE FIRAS and DIRBE data are due to
stellar and accretion disk radiation being downshifted into the
infrared by dust then redshifted into the submillimetre; they have
energy about twice all that in optical light, about a tenth of a
percent of that in the CMB. The spectrally well-defined
Sunyaev-Zeldovich (SZ) distortion associated with
Compton-upscattering of CMB photons from hot gas has not been
observed with FIRAS, but only at high resolution along
lines-of-sight through dozens of clusters
--- with very high signal-to-noise though. The FIRAS 95\% CL upper
limit of $ 6.0 \times 10^{-5}$ of the energy in the CMB is
compatible with the $\lta 10^{-5}$ expected from clusters, groups
and filaments in structure formation models, and places strong
constraints on the allowed amount of earlier energy injection, \eg
ruling out mostly hydrodynamic models of LSS.

{\bf Upper Limit Experiments from the 70s \& 80s:}  The story of
the experimental quest for anisotropies is a heroic
one.\footnote{Space constraints preclude adequate referencing
here, but these are given in \cite{bh95,lange00,jaffe00}.} The
original 1965 Penzias and Wilson discovery paper quoted angular
anisotropies below $10\%$, but by the late sixties $10^{-3}$
limits were reached, by Partridge and Wilkinson and by Conklin and
Bracewell. As calculations of baryon-dominated adiabatic and
isocurvature models improved in the 70s and early 80s, the
theoretical expectation was that the experimentalists just had to
get to $10^{-4}$, as they did, \eg Boynton and Partridge in 73.
The only signal found was the dipole, hinted at by Conklin and
Bracewell in 73, but found definitively in Berkeley and Princeton
balloon experiments in the late 70s, along with upper limits on
the quadrupole. Throughout the 1980s, the upper limits kept coming
down, punctuated by a few experiments widely used by theorists to
constrain models: the small angle 84 Uson and Wilkinson and 87
OVRO limits, the large angle 81 Melchiorri limit, early (87)
limits from the large angle Tenerife experiment, the small angle
RATAN-600 limits, the $7^\circ$-beam Relict-1 satellite limit of
87, and Lubin and Meinhold's 89 half-degree South Pole limit,
marking a first assault on the peak.

These upper limit experiments were highly useful, in particular to
rule out adiabatic baryon-dominated models. In the early 80s, dark
matter dominated universes lowered theoretical predictions by
about an order of magnitude. In the 84 to mid-90s period, many
groups developed codes to solve the perturbed Boltzmann--Einstein
equations when dark matter was present. Armed with these pre-COBE
computations, plus the LSS information of the time, a number of
very interesting models fell victim to the data: scale invariant
isocurvature cold dark matter models in 86, large regions of
parameter space for isocurvature baryon models in 87, inflation
models with radically broken scale invariance leading to enhanced
power on large scales in 87-89, CDM models with a decaying ($\sim
{\rm keV}$) neutrino if its lifetime was too long ($\gta 10 {\rm
yr}$) in 87 and 91. Also in this period there were some limited
constraints on "standard" CDM models, restricting $\Omega_{tot}$,
$\Omega_B$, and the amplitude parameter $\sigma_8$. ($\sigma_8^2$
is a bandpower for density fluctuations on a scale associated with
rare clusters of galaxies, $8\hmpc$, where ${\rm h}=H_0/(100 \kms
\mpc^{-1})$.)

{\bf Post-DMR Experiments:} The now familiar motley pattern of
anisotropies associated with $2 \le \ell \lta 20$ multipoles at
the $30 \mu K$ level revealed by COBE at $7^\circ$ resolution was
shortly followed by detections, and a few upper limits (UL), at
higher $\ell$ in 19 other ground-based (gb) or balloon-borne (bb)
experiments --- most with many fewer resolution elements than the
600 or so for COBE. Some predated in design and even data delivery
the 1992 COBE announcement. Proceeding from the period we began
analyzing them, we have the intermediate angle SP91 (gb), the
large angle FIRS (bb), both with strong hints of detection before
COBE, then, post-COBE,  more Tenerife (gb), MAX (bb), MSAM (bb),
white-dish (gb, UL), argo (bb), SP94 (gb), SK93-95 (gb), Python
(gb), BAM (bb), CAT (gb), OVRO-22 (gb), SuZIE (gb, UL), QMAP (bb),
VIPER (gb) and Python V (gb). A list valid to April 1999 with
associated bandpowers is given in \cite{bjk9800}, and are referred
here as 4.99 data. They showed evidence for a first peak
\cite{bjk9800}, although it was not well localized. Within limited
parameter sets, good constraints on $n_s$, some on $\Omega_{tot}$
and $\Omega_\Lambda$ could be given, when LSS was added.

{\bf The Present, TOCO, BOOMERANG \& MAXIMA:} The picture
dramatically improved this year, as results were announced first
in summer 99 from the ground-based TOCO experiment in Chile
\cite{toco98}, then in November 99 from the North American balloon
test flight of Boomerang \cite{mauskopf99}. These two additions
improved peak localization and gave evidence for $\Omega_{tot}\sim
1$. Then in April 2000 results  from the first CMB long duration
balloon (LDB) flight, Boomerang \cite{debernardis00}, were
announced,  followed in May 2000 by results from the night flight
of Maxima \cite{MAXIMA1}. Boomerang's best resolution was
$10^\prime$, about 40 times better than that of COBE, with tens of
thousands of resolution elements. Maxima had a similar resolution
but covered an order of magnitude less sky.

\begin{figure}[b!]
\vspace{-35pt} \centerline{\epsfig{file=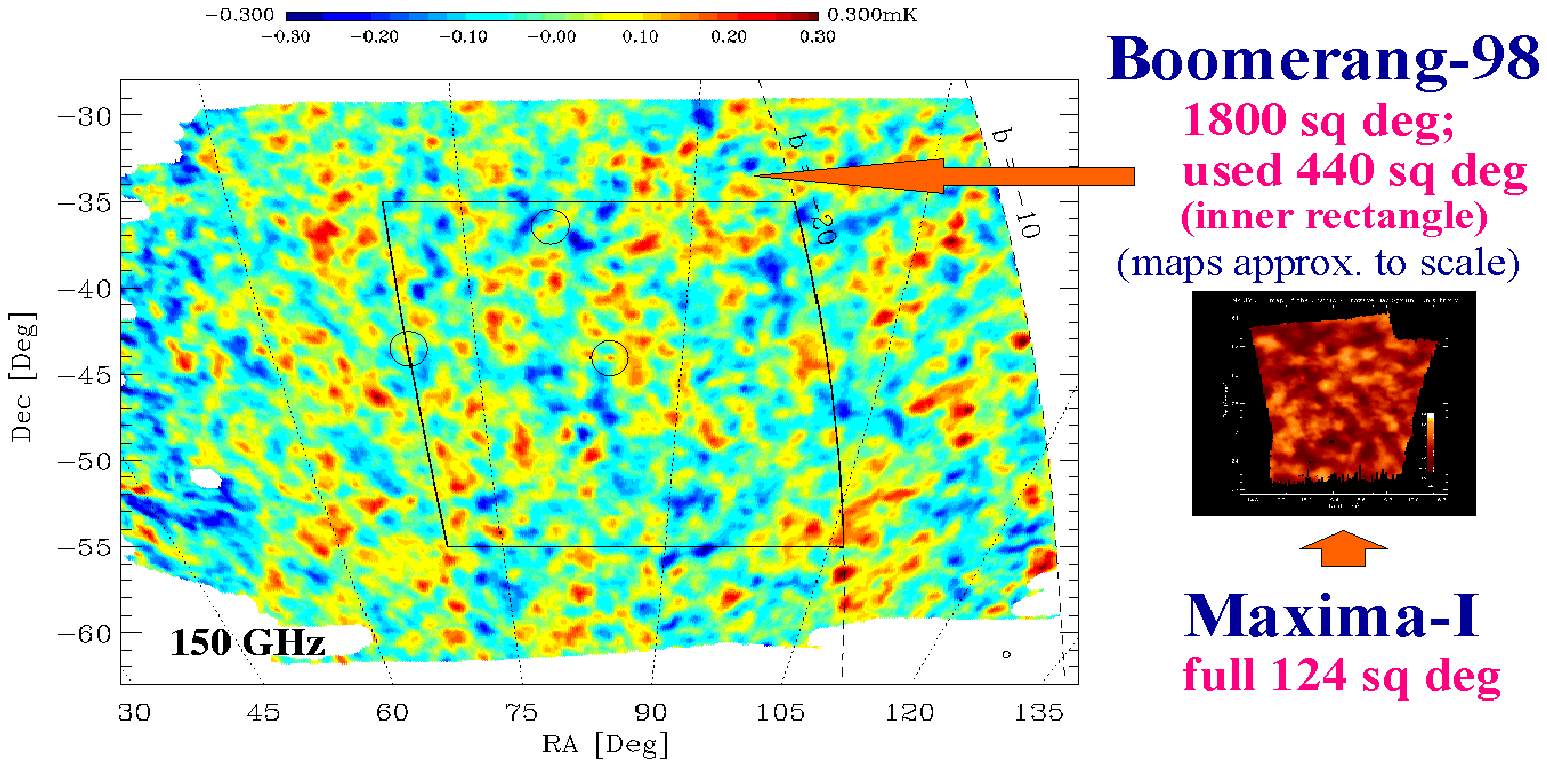,height=3.7in}}
\vspace{0pt}
\centerline{\epsfig{file=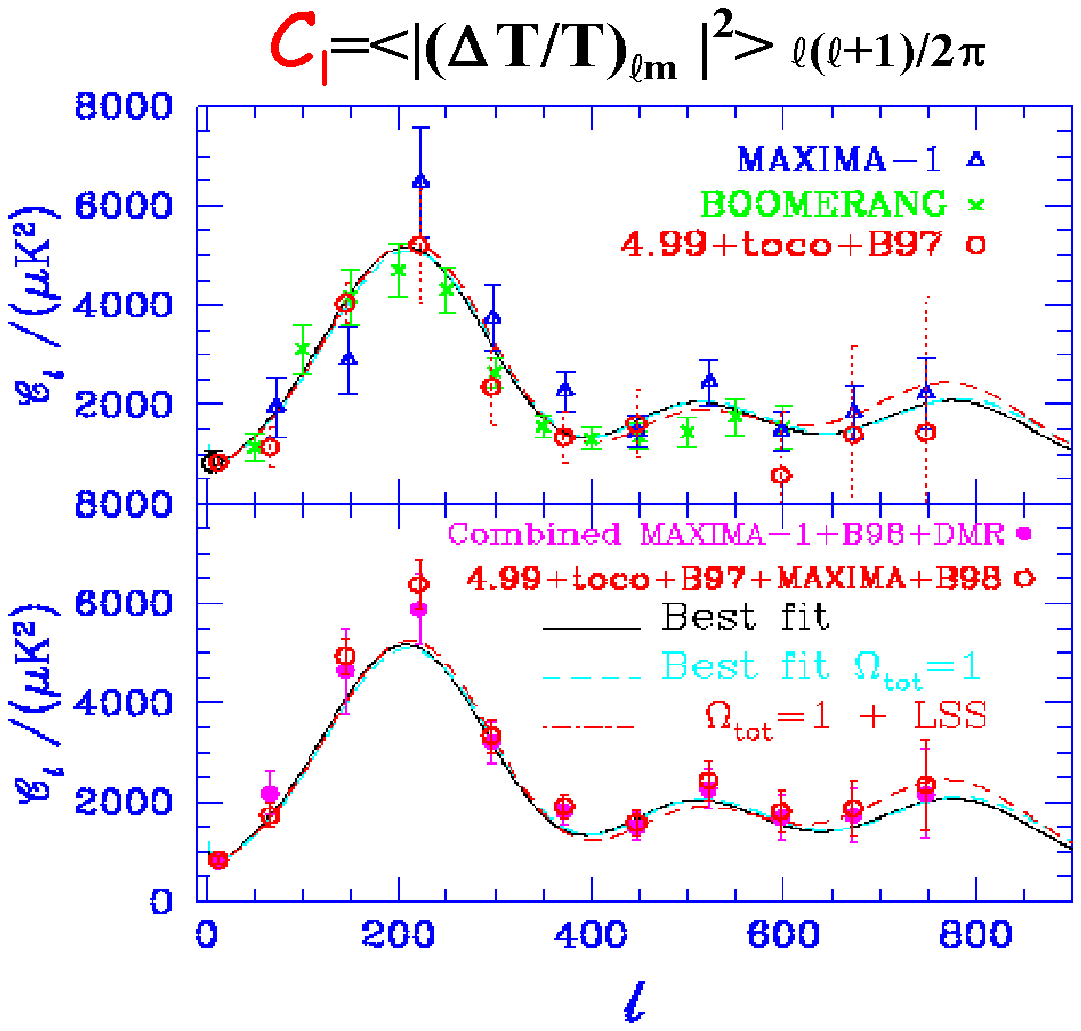,height=4.0in}}
\vspace{-5pt} \caption{\small The Boomerang 150A GHz bolometer map
(one out of 16) is shown in the top figure. Of the entire 1800
square degrees covered, only the interior 440 sq. degs. (within
the central rectangle) were used in our analysis, \ie in all less
than 5\% of the data.  The 124 square degree Maxima-I map, drawn
to scale, is also shown. In the lower figure, the ${\cal C}_\ell$
(defined in terms of CMB temperature anisotropy multipoles,
$(\Delta T/T)_{\ell m}$, as indicated) grouped in bandpowers for
prior-CMB experiments, including TOCO and the North American
Boomerang test flight, (squares)  are contrasted with the
Boomerang-LDB (crosses) and Maxima-I (triangles) results in the
upper panel. The lower panel shows the optimally-combined power
spectra of Boomerang+Maxima+prior-CMB (circles) and contrasts it
with that for Boomerang+Maxima+DMR (squares), showing that
including the prior experiments does not make a large difference
to the results. Best-fit models for arbitrary $\Omega_{tot}$ and
for $\Omega_{tot}$=1 are shown in both panels.\normalsize }
\label{fig:mapCLdat}
\end{figure}

Boomerang carried a  1.2m telescope with 16 bolometers cooled to
300 mK in the focal plane aloft from McMurdo Bay in Antarctica in
late December 1998, circled the Pole for 10.6 days and landed just
50 km from the launch site, only slightly damaged. In
\cite{debernardis00}, maps at 90, 150 and 220 GHz showed the same
spatial features and the intensities were shown to fall precisely
on the CMB blackbody curve. The fourth frequency channel at 400
GHz is dust-dominated. Fig.~\ref{fig:mapCLdat} shows the 150 GHz
map derived using only one of the 16 bolometers. Although
Boomerang altogether probed 1800 square degrees, only the region
in the rectangle covering 440 square degrees was used in the
analysis described in \cite{lange00,jaffe00} and this paper.
Fig.~\ref{fig:mapCLdat} also shows  the 124 square degree region
of the sky (in the Northern Hemisphere) that Maxima-1 probed.
Though Maxima was not an LDB, it did so well because its
bolometers were cooled even more than Boomerang's, to 100 mK,
leading to higher sensitivity per unit observing time, it had a
star camera so the pointing was well determined, and, further, all
frequency channels were used in creating its map.

{\bf Primary CMB Processes and Soundwave Maps at Decoupling:} Both
Boomerang and Maxima  were designed to measure the {\it primary}
anisotropies of the CMB, those which can be calculated using
linear perturbation theory. What we see in Fig.~\ref{fig:mapCLdat}
are, basically, two images of soundwave patterns that existed
about 300,000 years after the Big Bang, when the photons were
freed from the plasma. The visually evident structure on degree
scales is even more apparent in the power spectra of the Fourier
transform of the maps, which show a dominant (first acoustic)
peak, a less prominent (or non-existent) second one, and the
possible hint of a third one from Maxima. Fig.~\ref{fig:mapCLdat}
also shows that the quite heterogeneous 4.99+TOCO+Boomerang-NA mix
of CMB data is very consistent with what Boomerang-LDB and Maxima
show.

 The images are actually a projected mixture of dominant
and subdominant physical processes through the photon decoupling
"surface",  a fuzzy wall at redshift $z_r \sim 1100$, when the
Universe passed from optically thick to thin to Thomson scattering
over a comoving distance $\sim 10 \hmpc$. Prior to this, acoustic
wave patterns in the tightly-coupled photon-baryon fluid on scales
below the comoving "sound crossing distance" at decoupling, $\lta
100 \hmpc$ (\ie $\lta 100 \kpc$ physical), were viscously damped,
strongly so on scales below the $\sim 10 \hmpc$ thickness over
which decoupling occurred. After, photons freely-streamed along
geodesics to us, mapping (through the angular diameter distance
relation) the post-decoupling spatial structures in the
temperature to the angular patterns we observe now as the {\it
primary} CMB anisotropies. The maps are images projected through
the fuzzy decoupling surface of the acoustic waves (photon
bunching), the electron flow (Doppler effect) and the
gravitational potential peaks and troughs ("naive" Sachs-Wolfe
effect) back then. Free-streaming along our (linearly perturbed)
past light cone leaves the pattern largely unaffected, except that
temporal evolution in the gravitational potential wells as the
photons propagate through them leaves a further $\Delta T$
imprint, called the integrated Sachs-Wolfe effect. Intense
theoretical work over three decades has put accurate calculations
of this linear cosmological radiative transfer on a firm footing,
and there is a speedy, publicly available and widely used code for
evaluation of anisotropies in a variety of cosmological scenarios,
``CMBfast'' \cite{cmbfast}, including the latest hydrogen/helium
recombination evaluations, and  with extensions to more
cosmological models added by a variety of researchers.

Of course there are a number of nonlinear effects that are also
present in the maps. These {\it secondary} anisotropies include
weak-lensing by intervening mass, Thompson-scattering by the
nonlinear flowing gas once it became "reionized" at $z \sim 20$,
the thermal and kinematic SZ effects, and the red-shifted emission
from dusty galaxies. They all leave non-Gaussian imprints on the
CMB sky.

{\bf The Future, beyond 2000:} We are only at the beginning of the
high precision CMB era. HEMT-based interferometers are already in
place taking data: the VSA (Very Small Array) in Tenerife, the CBI
(Cosmic Background Imager) in Chile, DASI (Degree Angular Scale
Interferometer) at the South Pole, where the bolometer-based
single dish ACBAR experiment will operate this year. Other LDBs
will be flying within the next few years: Arkeops, Tophat,
Beast/Boost; and in 2001, Boomerang will fly again, this time
concentrating on polarization. As well, MAXIMA will fly as the
polarization-targeting MAXIPOL. In April 2001, NASA will launch
the all-sky HEMT-based MAP satellite, with $12^\prime$ resolution.
Further downstream, in 2007, ESA will launch the
bolometer+HEMT-based Planck satellite, with $5^\prime$ resolution.

Secondary anisotropies are also being targeted with new
instruments.  SZ anisotropies have been probed by single dishes,
the OVRO and BIMA mm arrays, and the Ryle interferometer. A number
of planned HEMT-based interferometers being built are more
ambitious: AMI (Britain), the JCA (Chicago), AMIBA (Taiwan), MINT
(Princeton). As well other kinds of bolometer-based experiments
will be used to probe the SZ effect, including the CSO (Caltech
submm observatory) with BOLOCAM on Mauna Kea, ACBAR at the South
Pole, the  LMT (large mm telescope) in Mexico, and the LDB BLAST.
Anisotropies from dust emission from high redshift galaxies are
being targeted by the JCMT with the SCUBA bolometer array, the
OVRO mm interferometer, the CSO,  the SMA (submm array) on Mauna
Kea, the LMT, the ambitious US/ESO ALMA mm array in Chile, the LDB
BLAST, and ESA's FIRST satellite. About $50\%$ of the submm
background has so far been identified with sources that SCUBA has
found.

{\bf The CMB Analysis Pipeline:} Analyzing Boomerang and other
experiments involves a pipeline that takes (1) the timestream in
each of the bolometer channels coming from the balloon plus
information on where it is pointing and turns it into (2) spatial
maps for each frequency characterized by average temperature
fluctuation  values in each pixel (Fig.~\ref{fig:mapCLdat}) and a
pixel-pixel correlation matrix characterizing the noise, from
which various statistical quantities are derived, in particular
(3) the temperature power spectrum as a function of multipole
(Fig.~\ref{fig:mapCLdat}), grouped into bands, and two band-band
error matrices which together determine the full likelihood
distribution of the bandpowers~\cite{bjk9800}. Fundamental to the
first step is the extraction of the sky signal from the noise,
using the only information we have, the pointing matrix mapping a
bit in time onto a pixel position on the sky.

There is generally another step in between (2) and (3), namely
separating the multifrequency spatial maps into the physical
components on the sky: the primary CMB, the thermal and kinematic
Sunyaev-Zeldovich effects, the dust, synchrotron and
bremsstrahlung Galactic signals, the extragalactic radio and
submillimetre sources. The strong agreement among the Boomerang
maps indicates that to first order we can ignore this step, but it
has to be taken into account as the precision increases.  The
Fig.~\ref{fig:mapCLdat} map is consistent with a Gaussian
distribution, thus fully characterized by just the power spectrum.
Higher order (concentration) statistics (3,4-point functions,
\etc) tell us of non-Gaussian aspects, necessarily expected from
the Galactic foreground and extragalactic source signals, but
possible even in the early Universe fluctuations. For example,
though non-Gaussianity occurs only in the more baroque inflation
models of quantum noise, it is a necessary outcome of
defect-driven models of structure formation. (Peaks compatible
with Fig.~\ref{fig:mapCLdat} do not appear in non-baroque defect
models, which now appear unlikely.) Though great strides have been
made in the analysis of Boomerang and Maxima, there is intense
needed effort worldwide now to develop new fast algorithms to deal
with the looming megapixel datasets of LDBs and the
satellites~\cite{bcjk99szapudi00}.

\section*{Cosmic Parameter Estimation}

{\bf Parameters of Structure Formation:} For this paper, we adopt
a restricted set of 8 cosmological parameters, augmenting the
basic 7 used in \cite{lange00,jaffe00},
$\{\Omega_\Lambda,\Omega_{k},\omega_b,\omega_{cdm}, n_s,\tau_C,
\sigma_8\}$, by one. The vacuum or dark energy encoded in the
cosmological constant $\Omega_\Lambda$ is reinterpreted as
$\Omega_Q$, the energy in a scalar field $Q$ which dominates at
late times, which, unlike $\Lambda$,  could have complex dynamics
associated with it. $Q$ is now often termed a quintessence field -
see http://feynman.princeton.edu/~steinh/ "Quintessence? - an
overview" for a pedagogical introduction.  One popular
phenomenology is to add one more parameter, $w_Q = p_Q /\rho_Q$,
where $p_Q$ and $\rho_Q$ are the pressure and density of the
$Q$-field, related to its kinetic and potential energy by $\rho_Q
= \dot{Q}^2/2+(\nabla Q)^2/2+V(Q)$, $ p_Q = \dot{Q}^2/2 -(\nabla
Q)^2/6-V(Q)$. Thus $w_Q=-1$ for the cosmological constant. Spatial
fluctuations of $Q$ are expected to leave a direct imprint on the
CMB for small $\ell$, typically smaller than Boomerang or Maxima
are sensitive to. We ignore this complication here. As well, as
long as $w_Q$ is not exactly $-1$, it will vary with time, but the
data will have to improve for there to be sensitivity to this, and
for now we can just interpret $w_Q$ as an appropriate time-average
of the equation of state. The curvature energy $\Omega_k \equiv
1-\Omega_{tot}$ also can dominate at late times, as well as
affecting the geometry.

We use only 2 parameters to characterize the early universe
primordial power spectrum of gravitational potential fluctuations
$\Phi$, one giving the overall power spectrum  amplitude ${\cal
P}_{\Phi}(k_n)$, and one defining the shape, a spectral tilt $n_s
(k_n) \equiv 1+d\ln {\cal P}_{\Phi}/d \ln k$, at some (comoving)
normalization wavenumber $k_n$. We really need another 2,  ${\cal
P}_{GW}(k_n)$ and $n_t(k_n)$, associated with the gravitational
wave component. In inflation, the amplitude ratio is related to
$n_t$ to lowest order, with ${\cal O}(n_s-n_t)$ corrections  at
higher order, \eg \cite{bh95}. There are also useful limiting
cases for the $n_s-n_t$  relation. However, as one allows the
baroqueness of the inflation models to increase, one can entertain
essentially any power spectrum (fully $k$-dependent $n_s(k)$ and
$n_t(k)$) if one is artful enough in designing inflaton potential
surfaces. As well, one can have more types of modes present, \eg
scalar isocurvature modes (${\cal P}_{is}(k_n),n_{is}(k)$) in
addition to, or in place of, the scalar curvature modes (${\cal
P}_{\Phi}(k_n),n_{s}(k)$). However, our philosophy is consider
minimal models first, then see how progressive relaxation of the
constraints on the inflation models, at the expense of increasing
baroqueness,  causes the parameter errors to open up. For example,
with COBE-DMR and Boomerang, we can probe the GW contribution, but
the data are not powerful enough to determine much. Planck can in
principle probe the gravity wave contribution reasonably well.

We use another 2 parameters to characterize the transport of the
radiation through the era of photon decoupling, which is sensitive
to the physical density of the various species of particles
present then, $\omega_j \equiv \Omega_j {\rm h}^2$. We really need
4: $\omega_b$ for the baryons, $\omega_{cdm}$ for the cold dark
matter, $\omega_{hdm}$ for the hot dark matter (massive but light
neutrinos), and $\omega_{er}$ for the relativistic particles
present at that time (photons, very light neutrinos, and possibly
weakly interacting products of late time particle decays). For
simplicity, though, we restrict ourselves to the conventional 3
species of  relativistic neutrinos plus photons, with
$\omega_{er}$ therefore fixed by the CMB temperature and the
relationship between the neutrino and photon temperatures
determined by the extra photon entropy accompanying $e^+ e^- $
annihilation. Of particular importance for the pattern of the
radiation is the (comoving) distance sound can have influenced by
recombination (at redshift $z_r= a_r^{-1}-1$), $r_s =
6000/\sqrt{3} \mpc \int_{0}^{\sqrt{a_r}} (\omega_m + \omega_{er}
a^{-1})^{-1/2} (1+ \omega_b a/(4\omega_\gamma /3))^{-1/2}\
d\sqrt{a}$, where $\omega_\gamma = 2.46 \times 10^{-5}$ is the
photon density, $\omega_{er} = 1.68 \omega_\gamma$ for 3 species
of massless neutrinos and $\omega_m \equiv
\omega_{hdm}+\omega_{cdm}+\omega_b$.

 The angular diameter distance relation,
 ${\cal R} =\{d_k {\rm sinh} (\chi_r/d_k), \chi_r, d_k {\rm sin} (\chi_r/d_k)\}$, where
$\chi_r = 6000 \mpc \int_{\sqrt{a_r}}^{1}  (\omega_m + \omega_Q
a^{-6w_Q} +\omega_k a)^{-1/2}\ d\sqrt{a}$ is the comoving distance
to recombination, $d_k =3000 |\omega_k|^{-1/2} \mpc$ is the
curvature scale and the 3 cases are for negative, zero and
positive mean curvature, adds dependence upon $\omega_{k}$,
$\omega_Q$ and $w_Q$ as well as on $\omega_m$. The location of the
first acoustic peak $L_{Pk}$ is proportional to the ratio of
${\cal R}$ to $r_s$, hence depends upon $\omega_b$ through the
sound speed as well.  Thus $L_{Pk}$ defines a functional
relationship among these parameters,  a {\it degeneracy}
\cite{degeneracies} that would be exact except for the integrated
Sachs-Wolfe effect, associated with the change of $\Phi$ with time
if $\Omega_Q$ or $\Omega_k$ is nonzero. (If $\dot{\Phi}$ vanishes,
the energy of photons coming into potential wells is the same as
that coming out, and there is no net impact of the rippled light
cone upon the observed $\Delta T$.)

 Our 7th parameter is an astrophysical one, the Compton "optical depth" $\tau_C$ from a
reionization redshift $z_{reh}$ to the present. It lowers ${\cal
C}_\ell$ by $\exp(-2\tau_C)$ at the high $\ell$'s probed by
Boomerang. For typical models of hierarchical structure formation,
we expect $\tau_C \lta 0.2$. It is partly degenerate with
$\sigma_8$ and cannot be determined at this precision by CMB data
now.

The LSS also depends upon our parameter set: the most important
combination is the wavenumber of the horizon when the energy
density in relativistic particles equals the energy density in
nonrelativistic particles: $k_{Heq}^{-1} \approx 5 \Gamma^{-1}
\hmpc$, where $\Gamma \approx \Omega_m {\rm h}
\Omega_{er}^{-1/2}$. Instead of ${\cal P}_\Phi (k_n)$ for the
amplitude parameter, we often use ${\cal C}_{10}$ at $\ell =10$
for CMB only, and $\sigma_8^2$ when LSS is added. When LSS is
considered in this paper, it refers to constraints on $\Gamma +
(n_s-1)/2$ and $\ln \sigma_8^2$ that are obtained by comparison
with the data on galaxy clustering and cluster abundances
\cite{lange00}.

When we allow for freedom in $\omega_{er}$, the abundance of
primordial helium, tilts of tilts ($dn_{\{s,is,t\}}(k_n)/d\ln k,
...$) for 3 types of perturbations, the parameter count would be
17, and many more if we open up full theoretical freedom in
spectral shapes. However, as we shall see, as of now only 3 or 4
combinations can be determined with 10\% accuracy with the CMB.
Thus choosing 8 is adequate for the present; 7 of these are
discretely sampled\cite{database}, with generous boundaries,
though for drawing cosmological conclusions we adopt a weak prior
probability on the Hubble parameter and age: we restrict ${\rm h}
$ to lie in the 0.45 to 0.9 range, and the age to be above 10 Gyr.

\begin{figure}[b!]
\vspace{-20pt}
\centerline{\hspace{15pt}\epsfxsize=3.5in\epsfbox{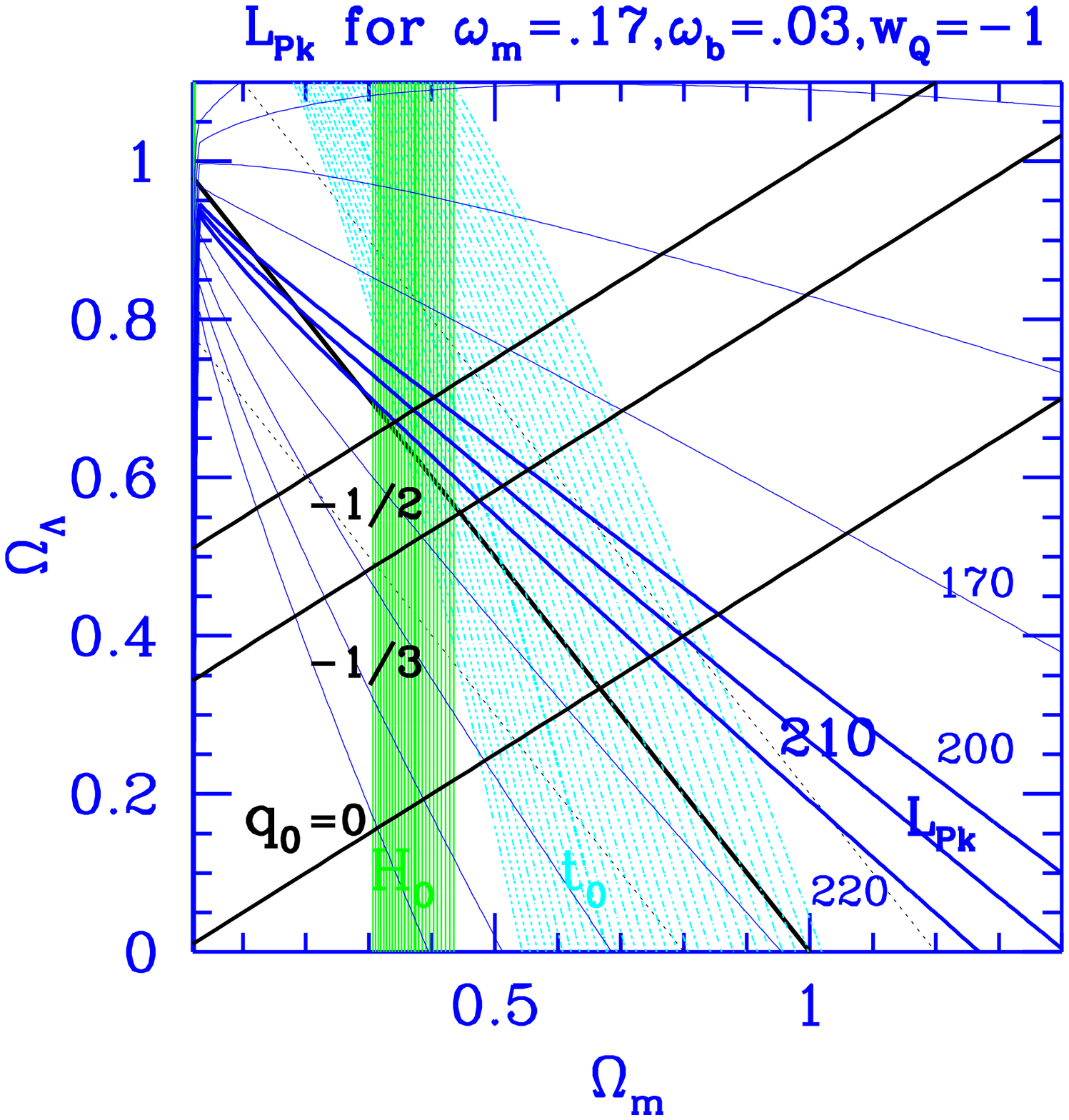}
\hspace{-25pt} \epsfxsize=5.0in\epsfbox{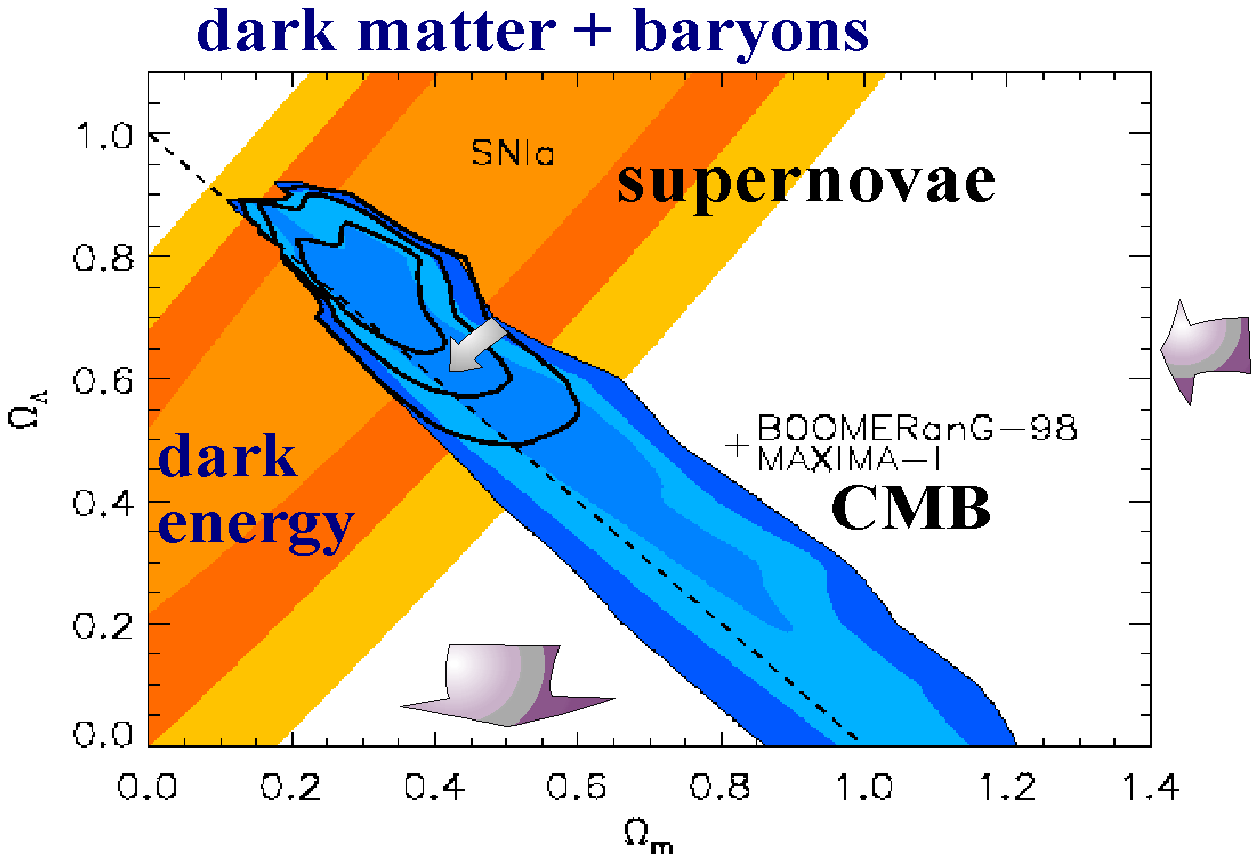}}
\vspace{-20pt}
\centerline{\hspace{15pt}\epsfxsize=3.5in\epsfbox{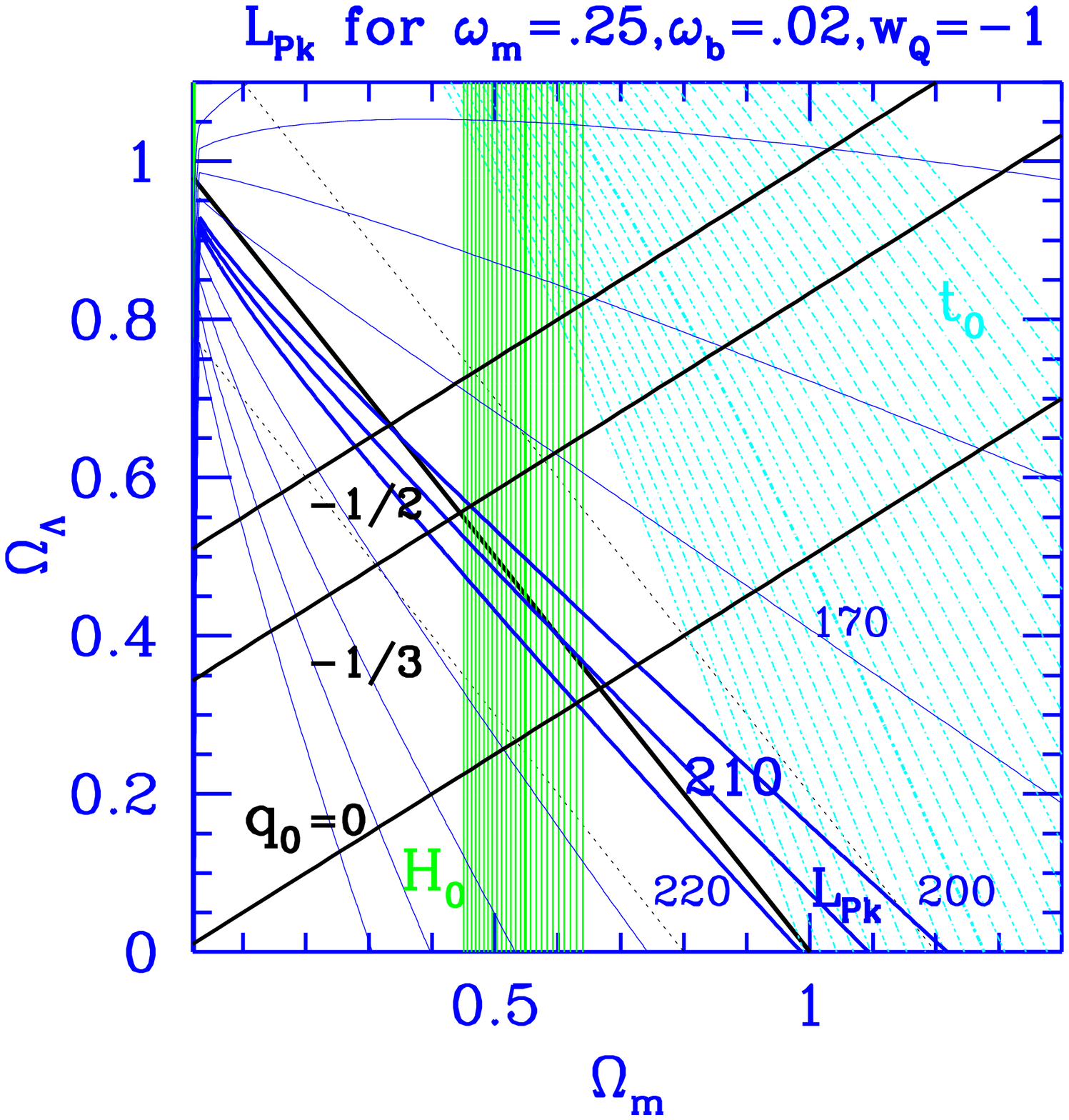}
 \hspace{-25pt} \epsfxsize=5.0in\epsfbox{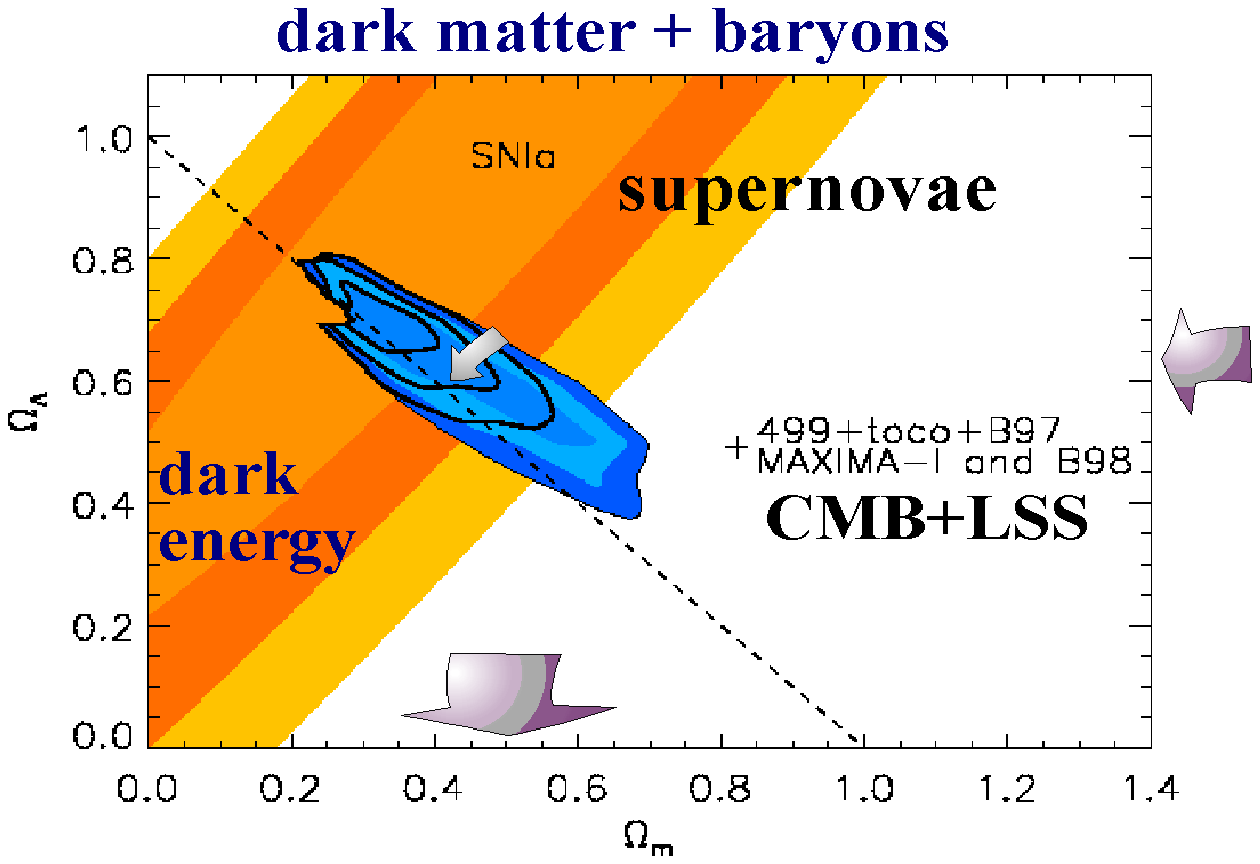}}
\vspace{-1pt} \caption{\small The left panels show lines of
constant $L_{Pk}$ in the $\Omega_m$--$\Omega_Q$ plane (assuming
$w_Q$=$-1$, \ie a cosmological constant) for two choices of $\{
\omega_m ,\omega_b \}$, the top the most probable values, the
bottom when the current BBN constraint is imposed (lowering
$\omega_b$ increases the sound speed, decreasing $L_{Pk}$, and
varying $\omega_m$ also shifts it). The $0.65 < {\rm h}< 0.75 $
(heavier shading, $H_0$) and $11< $ age $< 15$ (lighter shading,
$t_0$) ranges and decelerations $q_0=0,-1/3,-1/2$ are also noted.
The sweeping back of the $L_{Pk}$ curves into the closed models as
$\Omega_Q$ is lowered shows that even if $\Omega_{tot}$=1, the
phase space results in a 1D projection onto the $\Omega_{tot}$
axis that would be skewed to $\Omega_{tot}>1$, a situation we see
in Table~\ref{tab:exptparams}. The right panels show 1,2,3-sigma
likelihood contour shadings  for Boomerang+Maxima+DMR and the
weak-H+age prior probability (top left) and when LSS is added
(bottom left). The supernova contour shadings are also plotted,
and the solid contour lines are what you get when you combine the
two likelihoods.  In the bottom left panel, "prior-CMB"
experiments (including TOCO and Boomerang-NA), have been added as
well, although it makes little difference to the result whether
they are included or not. Note that the contours are near the
diagonal $\Omega_{tot}=1$ line, but also follow a weighted average
of $L_{Pk}\sim 210$ lines. This approximate degeneracy implies
$\Omega_Q$ is poorly constrained for CMB-only, but it is broken
when LSS is added, giving a solid SN1-independent $\Omega_Q$
"detection". \normalsize } \label{fig:OmOL}
\end{figure}

\begin{figure}[b!]
\vspace{-20pt}
\centerline{\hspace{5pt}\epsfxsize=3.5in\epsfbox{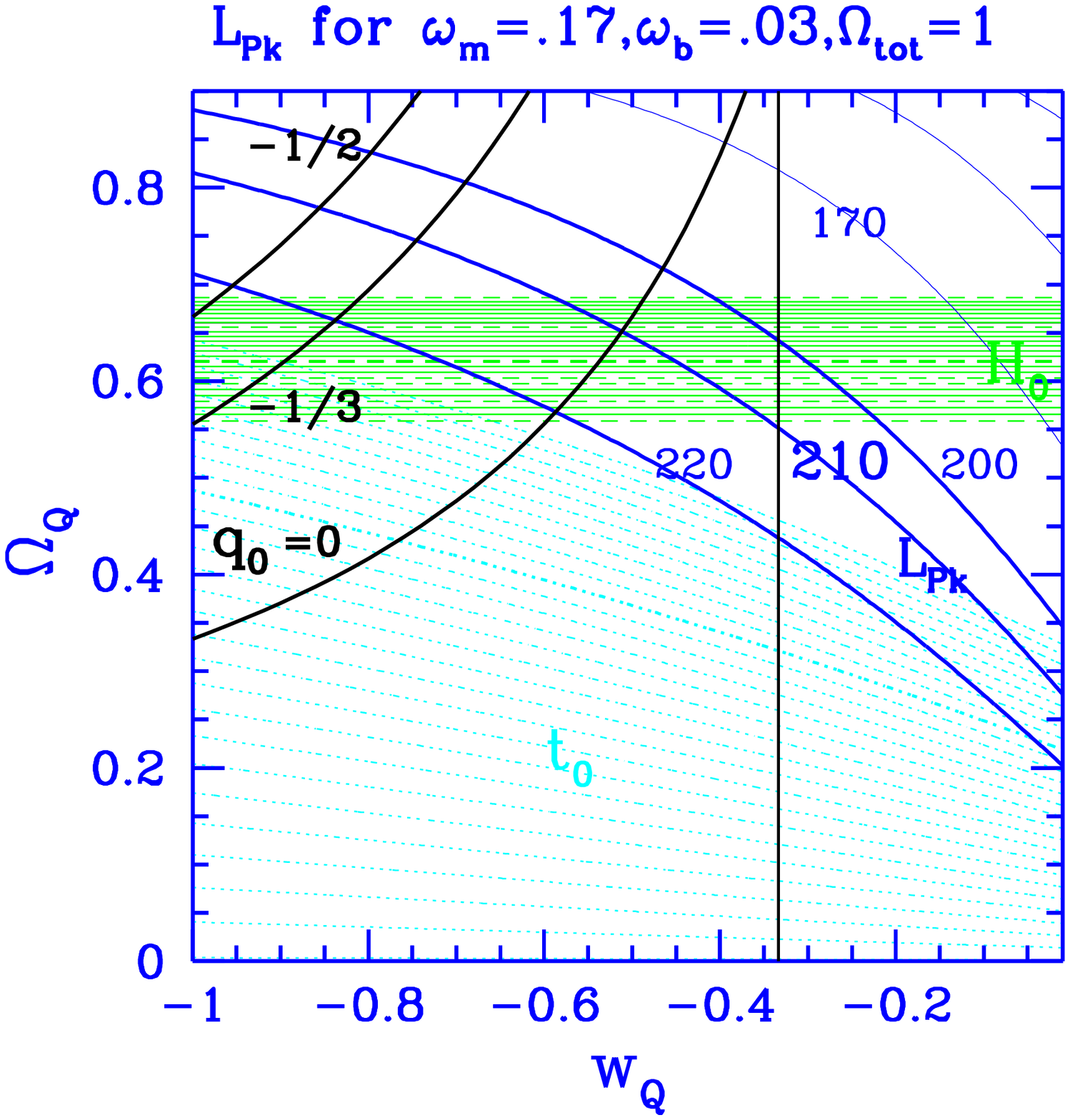}
\hspace{-15pt} \epsfxsize=3.5in\epsfbox{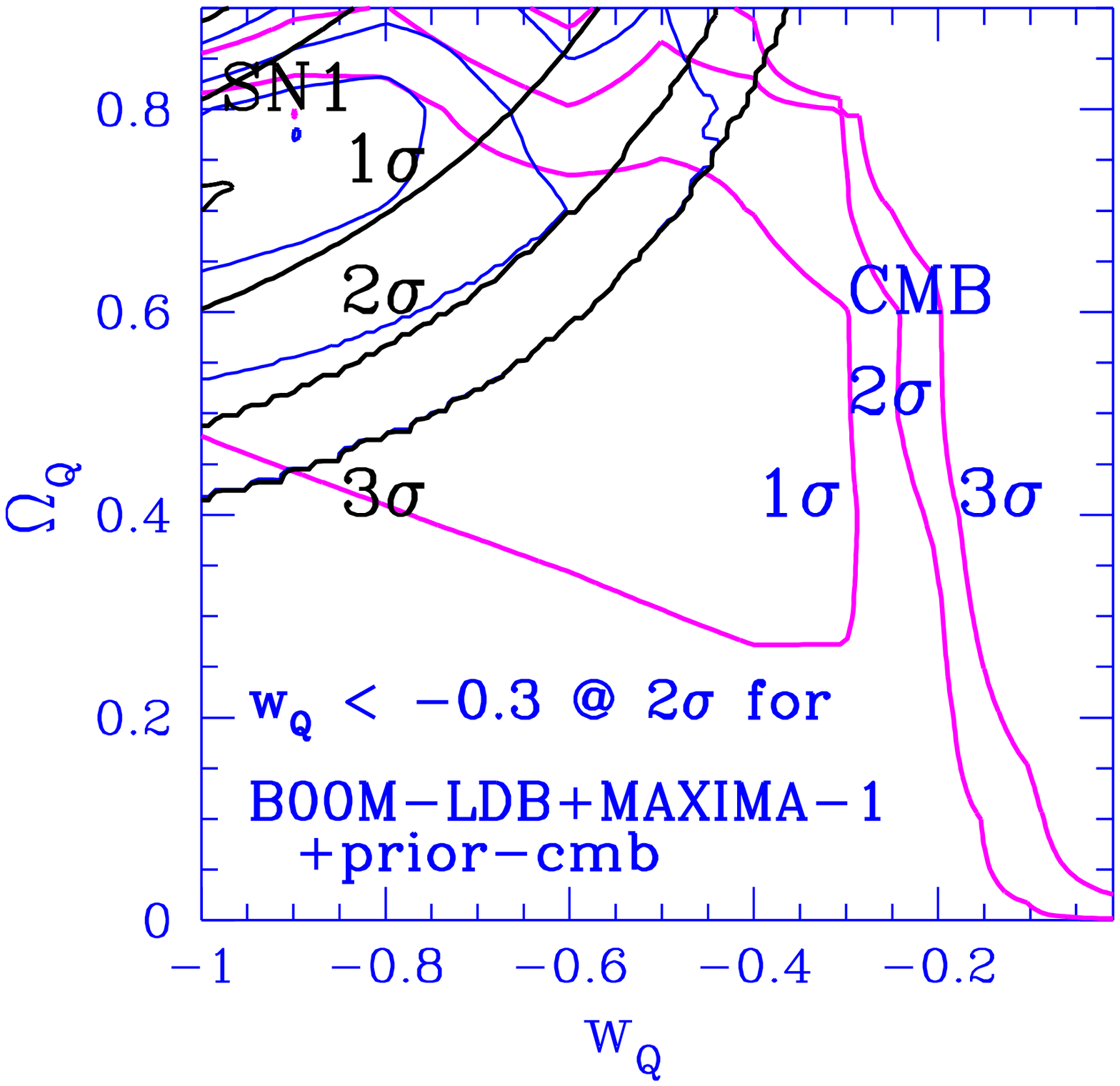}}
\vspace{-20pt}
\centerline{\hspace{5pt}\epsfxsize=3.5in\epsfbox{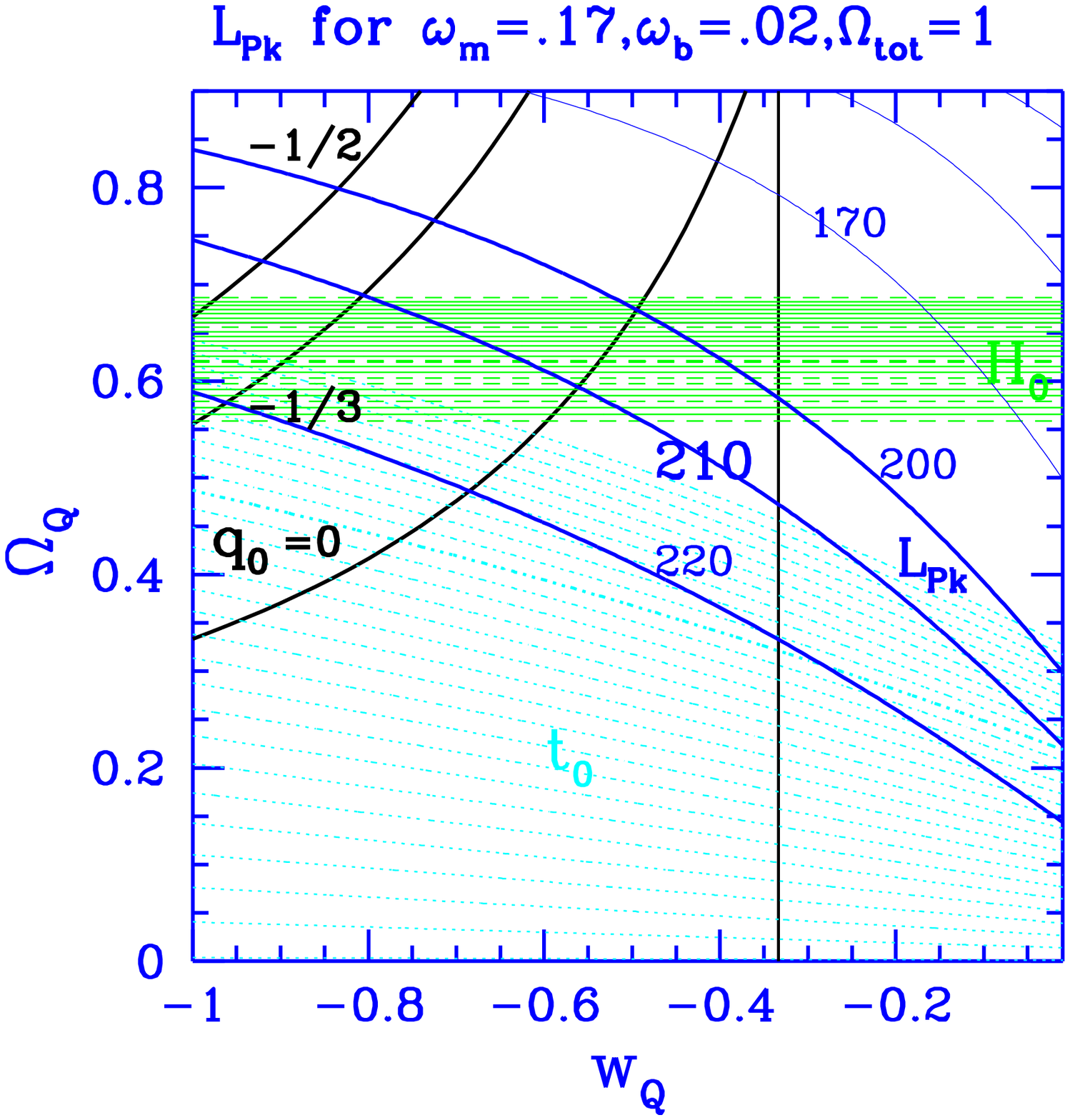}
\hspace{-15pt} \epsfxsize=3.5in\epsfbox{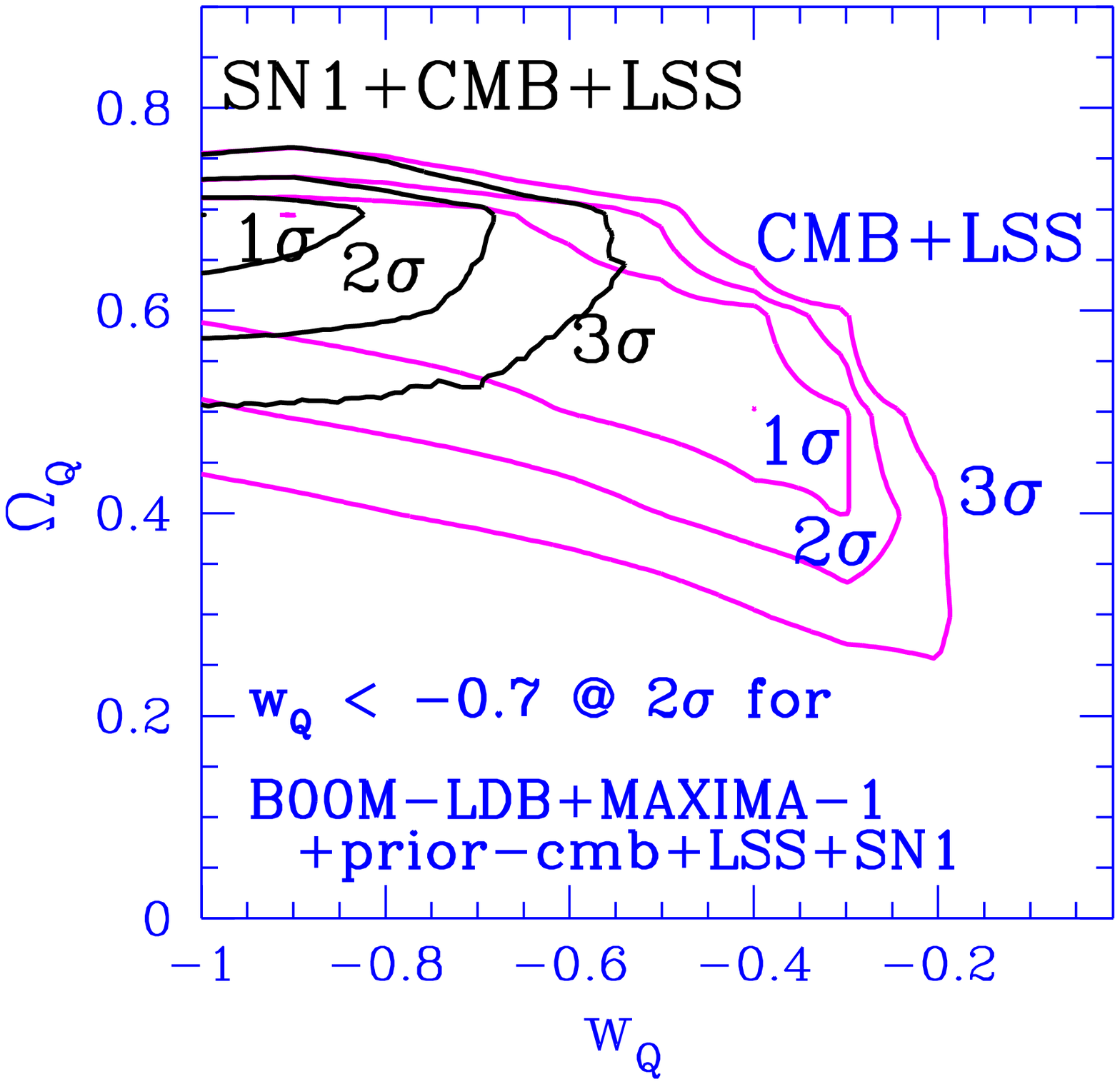}}
\vspace{-5pt} \caption{\small In the left panels, lines of
constant $L_{Pk}$ in the $w_Q$--$\Omega_Q$ quintessence plane
(with $\Omega_{tot}$=1) are shown for the most probable values of
$\{ \omega_m ,\omega_b \}$, and when $\omega_b$ is constrained to
the BBN value of 0.02. Lines of constant deceleration parameter
$q_0 = (\Omega_m +(1+3w_Q)\Omega_Q )/2$ and $0.65 < {\rm h}< 0.75
$ (heavier shading, $H_0$) and $11< $ age $< 15$ (light shading,
$t_0$) ranges are also shown. The right panels show 1,2,3-sigma
likelihood contours for Boomerang+Maxima+prior-CMB data with the
weak-H+age prior probability and $\Omega_{tot}$=1. Top shows CMB
only, bottom CMB+LSS. SN1 contours,  provided by Saul
Perlmutter~\protect\cite{perlmutter99}, and the CMB+SN1 combined
ones are also shown in the top, and the CMB+LSS+SN1 ones in the
bottom. Thus, little follows from CMB only, $\Omega_Q$ gets
localized when LSS is added but not $w_Q$, and $w_Q$ localization
with SN1 is mainly because of SN1. The 1,2,3-sigma lines for SN1
shown in the upper right are rather similar to the constant
deceleration parameter lines in the left panels. \normalsize }
\label{fig:wQOQ}
\end{figure}

{\bf The First Peak and $\Omega_{tot}$, $\Omega_Q$ and $w_Q$:} For
given $\omega_m$ and $\omega_b$,  we show the lines of constant
$L_{Pk}\propto {\cal R}/r_s$ in the $\Omega_{tot}$--$\Omega_Q$
plane for $w_Q$=$-1$ in Fig.~\ref{fig:OmOL}, and in the
$w_Q$--$\Omega_Q$ plane for $\Omega_{tot}$=1 in
Fig.~\ref{fig:wQOQ}, using the formulas given above and in
\cite{degeneracies}.  Our current best estimate~\cite{bnu2K} of
$L_{Pk}$, using all current CMB data, is $212\pm 7$, obtained by
forming $\exp<\ln L_{Pk}>$, where the average and variance of $\ln
L_{Pk}$ are determined by integrating over the
probability-weighted database described above, restricted here to
the $\tau_C=0$ part. With just the prior-CMB data the value was
$224\pm 25$, showing how it has localized.  The numbers change a
bit depending upon exactly what database or functional forms one
averages over. The constant $L_{Pk}$ lines look rather similar to
the contours shown in the right panel, showing that the ${\cal
R}/r_s$ degeneracy plays a large role in determining the contours.
The contours hug the $\Omega_{tot}=1$ line more closely than the
allowed $L_{Pk}$ band does for the maximum probability values of
$\omega_m$ and $\omega_b$, because of the shift in the allowed
$L_{Pk}$ band as $\omega_m$ and $\omega_b$ vary in this plane.

{\bf Marginalized Estimates of our Basic 8 Parameters:}
Table~\ref{tab:exptparams} shows there are strong detections with
only the CMB data for $\Omega_{tot}$, $\omega_b$ and $n_s$ in the
minimal inflation-based 8 parameter set. The ranges quoted are
Bayesian 50\% values and the errors are 1-sigma, obtained after
projecting (marginalizing) over all other parameters. With Maxima,
$\omega_{cdm}$ begins to localize, but much more so when LSS
information is added. Indeed, even with just the COBE-DMR+LSS
data, $\omega_{cdm}$ is already localized. That $\Omega_Q$ is not
well determined is a manifestation of the
$\Omega_{tot}$--$\Omega_Q$ near-degeneracy discussed above, which
is broken when LSS is added because the CMB-normalized $\sigma_8$
is quite different for open cf. pure $Q$-models. Supernova at high
redshift give complementary information to the CMB, but with
CMB+LSS (and the inflation-based paradigm) we do not need it: the
CMB+SN1 and CMB+LSS numbers are quite compatible. In our space,
the Hubble parameter, ${\rm h}= (\sum_j (\Omega_j{\rm h}^2
))^{1/2}$, and the age of the Universe, $t_0$, are derived
functions of the $\Omega_j{\rm h}^2$: representative values are
given in the Table caption. CMB+LSS does not currently give a
useful constraint on $w_Q$, though $w_Q \lta -0.7$ with SN1.

\begin{table}
\caption{Cosmological parameter values and their 1-sigma errors
are shown, determined after marginalizing over the other $6$
cosmological and $4^{+}$ experimental parameters, for
B98+Maxima-I+prior-CMB and the weak prior used in
\protect\cite{lange00,jaffe00} ($0.45 \le {\rm h} \le 0.9$, age $>
10$ Gyr). The LSS prior was also designed to be weak. The
detections are clearly very stable if extra "prior" probabilities
for LSS and SN1 are included. (Indeed, they are stable to
inclusion of stronger priors --- except if the BBN-derived $0.019
\pm 0.002$ is imposed \protect\cite{lange00}.) Similar tables for
B98+DMR are given in \protect\cite{lange00} and for
B98+MAXIMA-I+DMR in \protect\cite{jaffe00}. If $\Omega_{tot}$ is
varied, but $w_Q=-1$, parameters derived from our basic 8 come out
to be: age=$13.2\pm 1.3$ Gyr, ${\rm h}=0.70 \pm 0.09$,
$\Omega_m=0.35\pm .06$, $\Omega_b=0.065 \pm .02$. Restriction to
$\Omega_{tot}=1$ and $w_Q=-1$ yields: age=$11.6\pm 0.4$ Gyr, ${\rm
h}=0.80\pm .04$, $\Omega_m=0.31\pm .03$, $\Omega_b=0.05 \pm .005$;
allowing $w_Q$ to vary yields quite similar results. }
\label{tab:exptparams}
\begin{center}
\begin{tabular}{|l|llll|}
\hline
   & cmb & +LSS & +SN1 &  +SN1+LSS \\
\hline
 & $\Omega_{tot}$ & variable & $w_Q=-1$ & CASE  \\
\hline $\Omega_{tot}$           & $1.09^{+.07}_{-.07}$ &
$1.08^{+.06}_{-.06}$ & $1.04^{+.06}_{-.05}$ & $1.04^{+.05}_{-.04}$ \\
$\Omega_b{\rm h}^2$             & $.031^{+.005}_{-.005}$ &
$.031^{+.005}_{-.005}$ & $.031^{+.005}_{-.005}$ &
$.031^{+.005}_{-.005}$  \\
$\Omega_{cdm}{\rm h}^2$ & $.17^{+.06}_{-.05}$ &
$.14^{+.03}_{-.02}$ & $.13^{+.05}_{-.05}$ & $.15^{+.03}_{-.02}$  \\
$n_s$            & $1.05^{+.09}_{-.08}$ & $1.04^{+.09}_{-.08}$ &
$1.05^{+.10}_{-.09}$ & $1.06^{+.08}_{-.08}$ \\
$\Omega_{Q}$ & $0.48^{+.20}_{-.26}$ & $0.63^{+.08}_{-.09}$ &
$0.72^{+.07}_{-.07}$ & $0.70^{+.04}_{-.05}$  \\
\hline
 & $\Omega_{tot}$  & =1 & $w_Q=-1$ & CASE  \\
\hline $\Omega_b{\rm h}^2$& $.030^{+.004}_{-.004}$ &
$.030^{+.003}_{-.004}$ & $.030^{+.004}_{-.004}$ &
$.030^{+.003}_{-.004}$ \\
$\Omega_{cdm}{\rm h}^2$& $.19^{+.06}_{-.05}$ & $.17^{+.02}_{-.02}$
& $.16^{+.03}_{-.03}$ & $.17^{+.01}_{-.02}$\\
$n_s$ & $1.02^{+.08}_{-.07}$ & $1.03^{+.08}_{-.07}$ &
$1.03^{+.08}_{-.07}$ & $1.04^{+.07}_{-.07}$  \\
$\Omega_{Q}$& $0.58^{+.17}_{-.27}$ & $0.66^{+.04}_{-.06}$ &
$0.71^{+.06}_{-.07}$ & $0.69^{+.03}_{-.05}$ \\
\hline
 & $\Omega_{tot}$  & =1 &  $w_Q$ variable & CASE  \\
\hline $\Omega_b{\rm h}^2$& $.030^{+.004}_{-.004}$ &
$.030^{+.004}_{-.004}$ & $.030^{+.004}_{-.004}$ &
$.030^{+.004}_{-.004}$ \\
$\Omega_{cdm}{\rm h}^2$& $.17^{+.06}_{-.05}$ & $.16^{+.02}_{-.03}$
& $.14^{+.04}_{-.03}$ & $.17^{+.01}_{-.02}$\\
$n_s$ & $1.01^{+.08}_{-.07}$ & $1.02^{+.07}_{-.06}$ &
$1.01^{+.07}_{-.07}$ & $1.03^{+.07}_{-.06}$  \\
$\Omega_{Q}$& $0.56^{+.17}_{-.25}$ & $0.59^{+.08}_{-.10}$ &
$0.74^{+.06}_{-.08}$ & $0.68^{+.03}_{-.05}$ \\
$w_{Q}$ (95\%)& $< -0.29$ & $<-0.33$ & $< -0.69$ & $<-0.73$  \\
\hline
\end{tabular}
\end{center}
\end{table}

{\bf The Influence of Light Massive Neutrinos:} In \cite{bnu2K},
we considered what happens as we let $\Omega_{m\nu}/\Omega_m$, the
fraction of the matter in massive neutrinos,  vary from 0 to 0.3,
for Boomerang+Maxima+prior-CMB+LSS when the weak-H+age +
$\Omega_{tot}=1$ prior probability is adopted. Until Planck
precision, the CMB data by itself will not be able to strongly
discriminate this ratio. Adding HDM does have a strong impact on
the CMB-normalized $\sigma_8$ and the shape of the density power
spectrum (effective $\Gamma$ parameter), both of which mean that
when LSS is included, adding some HDM to CDM is strongly preferred
in the absence of $\Omega_Q$. However, though more (cold+hot) dark
matter is preferred at the expense of less dark energy,
significant $\Omega_Q$ is still required~\cite{mnu}. The
$\omega_b$ and $n_s$ likelihood curves are essentially independent
of $\Omega_{m\nu}/\Omega_m$.

{\bf The Future, Forecasts for Parameter Eigenmodes:}   We can
also forecast dramatically improved precision with further
analysis of Boomerang and Maxima, future LDBs, MAP and Planck.
Because there are correlations among the physical variables we
wish to determine, including a number of near-degeneracies beyond
that for $\Omega_{tot}$--$\Omega_Q$ \cite{degeneracies}, it is
useful to disentangle them, by making combinations which
diagonalize the error correlation matrix, "parameter eigenmodes"
\cite{bh95,degeneracies}.  For this exercise, we will add
$\omega_{hdm}$ and $n_t$ to our parameter mix, but set $w_Q$=$-1$,
making 9. (The ratio ${\cal P}_{GW}(k_n)/{\cal P}_\Phi (k_n)$ is
treated as fixed by $n_t$, a reasonably accurate inflation theory
result.) The forecast for Boomerang based on the 440 sq. deg.
patch with a single 150 GHz bolometer used in the published data
is 3 out of 9 linear combinations should be determined to $\pm
0.1$ accuracy. This is indeed what we get in the full analysis CMB
only for Boomerang+DMR. If 4 of the 6 150 GHz channels are used
and the region is doubled in size, we predict 4/9 could be
determined to $\pm 0.1$ accuracy. The Boomerang team is still
working on the data to realize this promise. And if the optimistic
case for all the proposed LDBs is assumed, 6/9 parameter
combinations could be determined to $\pm 0.1$ accuracy, 2/9 to
$\pm 0.01$ accuracy. The situation improves for the satellite
experiments: for MAP, we forecast  6/9 combos to $\pm 0.1$
accuracy, 3/9 to $\pm 0.01$ accuracy; for Planck,  7/9 to $\pm
0.1$ accuracy, 5/9 to $\pm 0.01$ accuracy. While we can expect
systematic errors to loom as the real arbiter of accuracy, the
clear forecast is for a very rosy decade of high precision CMB
cosmology that we are now fully into.

\normalsize


\begin{references}
\bibitem{mnu} \small The simplest interpretation of the
superKamiokande data on atmospheric $\nu_\mu$ is that
$\Omega_{\nu_\tau} \sim 0.001$, about the energy density of stars
in the universe, which implies a cosmologically negligible effect.
Degeneracy between \eg $\nu_\mu$ and $\nu_\tau$ could lead to the
cosmologically very interesting $\Omega_{m\nu}\equiv \Omega_{hdm}
\sim .1$, although the coincidence of closely related energy
densities for baryons, CDM, HDM and dark energy required would be
amazing. \normalsize
\bibitem{database} \small The specific discrete parameter values used
for the ${\cal C}_\ell$-database in this analysis were: ($\Omega_Q
=$ 0,.1,.2,.3,.4,.5,.6,.7,.8,.9,1.0,1.1), ($\Omega_k =$
.9,.7,.5,.3,.2,.15,.1,.05,0,-.05,-.1,-.15,-.2,-.3,-.5) \& ($\tau_c
=$0, .025, .05, .075, .1, .15, .2, .3, .5) when $w_Q=$ --1;
($\Omega_Q =$ 0,.1,.2,.3,.4,.5,.6,.7,.8,.9), ($w_Q=$
-1,-.9,-.8,-.7,-.6,-.5,-.4,-.3,-.2,-.1,-.01) \& ($\tau_c =$0,
.025, .05, .075, .1, .15, .2, .3, .5) when $\Omega_k$=0. For both
cases,  ($\omega_c =$ .03, .06, .12, .17, .22, .27, .33, .40, .55,
.8), ($\omega_b =$ .003125, .00625, .0125, .0175, .020, .025,
.030, .035, .04, .05, .075, .10, .15, .2), ($n_s = $1.5, 1.45,
1.4, 1.35, 1.3, 1.25, 1.2, 1.175, 1.15, 1.125, 1.1, 1.075, 1.05,
1.025, 1.0, .975, .95, .925, .9, .875, .85, .825, .8, .775, .75,
.725, .7, .65, .6, .55, .5), $\sigma_8^2$ was continuous, and
there were 4 experimental parameters, calibration and beam
uncertainties, for Boomerang and Maxima, as well as other
calibration parameters for some of the prior-CMB experiments.
\normalsize
\end{references}
\end{document}